\documentclass[12pt]{article}
\usepackage{graphicx}
\usepackage{multirow}

\usepackage[T1]{fontenc}
\usepackage[sc]{mathpazo}
\usepackage{amsmath}
\usepackage{amssymb}
\usepackage{enumerate}
\usepackage{amsthm}
\usepackage{amsfonts,mathrsfs}

\usepackage{hyperref}
\hypersetup{pdfpagemode=UseNone}

\newcommand{\tinyspace}{\mspace{1mu}}

\newcommand{\abs}[1]{\left\lvert\tinyspace #1 \tinyspace\right\rvert}

\newenvironment{mylist}[1]{\begin{list}{}{
    \setlength{\leftmargin}{#1}
    \setlength{\rightmargin}{0mm}
    \setlength{\labelsep}{2mm}
    \setlength{\labelwidth}{8mm}
    \setlength{\itemsep}{0mm}}}
    {\end{list}}

\newcommand{\Pa}[1]{\left(#1\right)}

\newcommand{\Br}[1]{\left[#1\right]}
\newcommand{\set}[1]{\{#1\}}
\newcommand{\Set}[1]{\left\{#1\right\}}

\newcommand{\fontmapset}{\mathbf} 

\newcommand{\Mset}[2]{\fontmapset{#1}\Pa{#2}}
\newcommand{\Lin}[1]{\Mset{L}{#1}}

\theoremstyle{definition}

\numberwithin{equation}{section}

\newcounter{questionnumber}

\begin{document}

\title{\Large\bf Analytical approximation for Landau's constants by using BPES method}

\author{Karem Boubaker\footnote{E-mail:mmbb11112000@yahoo.fr}\\[1mm]
{\it\small Ecole Superieure des Sciences et Techniques de Tunis , Universit\'{e} de Tunis, Tunis, Tunisia}
\\Lin Zhang\footnote{E-mail: godyalin@163.com}\\[1mm]
{\it\small Institute of Mathematics, Hangzhou Dianzi University,
Hangzhou 310018, P.R.~China}}

\date{}
\maketitle \mbox{}\hrule\mbox\\
\begin{abstract}

In this study, approximation formulas for evaluating Landau
constants are elaborated by using the Boubaker Polynomials Expansion
Scheme (BPES). Results are compared to some referred studies.\\~\\
\textbf{Keywords:} Landau constants; Approximation  theory; Complex
analysis; Boubaker Polynomials Expansion Scheme (BPES); Error
analysis.
\end{abstract}
\mbox{}\hrule\mbox\\~\\

\section{Introduction}

Landau's constants are defined \cite{Landau}-\cite{Zhao} for all
positive integers $n$, by:
\begin{eqnarray}\label{eq-1}
G_n = \sum^n_{k=0} \frac1{16^k}\binom{2k}{k}^2 = 1 +
\Pa{\frac{1}{2}}^2 + \Pa{\frac{1\cdot 3}{2\cdot 4}}^2 +\cdots +
\Pa{\frac{(2n-1)!!}{(2n)!!}}^2.
\end{eqnarray}
These constants were defined in relation with  a set $F$ of complex
analytic functions $f$ defined on an open region containing the
closure of the unit disk $D$ ($D = \set{z:\abs{z}<1}$) satisfying
the conditions:
\begin{eqnarray}\label{eq-2}
f(z)|_{z=0}=0\quad\text{and}\quad \frac{d}{dz}f(z)|_{z=0}=1.
\end{eqnarray}
If $\ell(f)$ is the supremum of all numbers such that $f(D)$
contains a disk of radius $1$, and that $f$ verifies the additional
conditions:
\begin{eqnarray}\label{eq-3}
\abs{f(z)}<1(|z|<1) \quad\text{and}\quad  f(z)=\sum^\infty_{k=0}a_k
z^k
\end{eqnarray}
then:
\begin{eqnarray}\label{eq-4}
L = \inf\Set{\ell(f): f\in F} = \abs{\sum^\infty_{k=0}a_k}\leqslant
G_n
\end{eqnarray}
In this paper, a convergent protocol is proposed in order to give
analytical expressions to the Landau's constants.

\section{Resolution process}

In concordance of the approximations performed by Watson
\cite{Watson}, Zhao \cite{Zhao} and Popa \cite{Popa}, a sequence
$\set{\omega_n}_{n\geqslant0}$ is defined as:
\begin{eqnarray}\label{eq-5}
\omega_n = G_n - \frac1{\pi} \ln(n+A) - \frac1{\pi}(\gamma + \ln 16)
+ \frac{B}{n+C},
\end{eqnarray}
where $\gamma$ is Euler's constant and $A,B$ and $C$ are unknown
constants.

The resolution process aims to find accurate approximation to the
values of the parameters $A,B$ and $C$ such that
$\set{\omega_n}_{n\geqslant0}$ is the fastest sequence which would
converge to zero.

\section{Approximation using The Boubaker Polynomials Expansion Scheme (BPES)}

\subsection{Presentation}

The Boubaker Polynomials Expansion Scheme BPES
\cite{Milgram}-\cite{Rahmanov} is a resolution protocol which has
been successfully applied to several applied-physics and mathematics
problems. The BPES protocol ensures the validity of the related
boundary conditions regardless main equation features. The BPES is
mainly based on Boubaker polynomials first derivatives properties:

\begin{eqnarray}\label{eq-6}
\sum^N_{k=1}B_{4k}(x)\big|_{x=0} = -2N\neq0,\quad
\sum^N_{k=1}B_{4k}(x)\big|_{x=r_k}=0
\end{eqnarray}
and
\begin{eqnarray}\label{eq-7}
\sum^N_{k=1}\frac{dB_{4k}(x)}{dx}\big|_{x=0} = 0 ,\quad
\sum^N_{k=1}\frac{dB_{4k}(x)}{dx}\big|_{x=0} = \sum^N_{k=1}H_k
\end{eqnarray}
with
$$
H_n = B_{4n}(r_n) =
\frac{4\alpha_n[2-r^2_n]\sum^n_{k=1}B^2_{4k}(r_n)}{B_{4(n+1)}(r_n)}
+4r^3_n.
$$
Several solution have been proposed through the BPES in
many fields such as numerical analysis \cite{Milgram}-\cite{Slama2},
theoretical physics \cite{Lazzez}-\cite{Fridjine1}, mathematical
algorithms \cite{Khelia}, heat transfer \cite{Mahmoud}, homodynamic
\cite{Dada,Tabatabaei}, material characterization \cite{Belhadj1},
fuzzy systems modeling \cite{Belhadj2}-\cite{Fridjine2} and biology
\cite{Benhaliliba,Rahmanov}.

\subsection{Application}

 The resolution protocol is based on setting
 $\widehat{A},\widehat{B}$ and $\widehat C$ as estimators to the
 constants $A,B$ and $C$, respectively:
\begin{eqnarray}
\left\{\begin{array}{ccc}
\widehat A & = & \frac1{2N_0}\sum^{N_0}_{k=1}\xi^A_k B_{4k}(xr_k), \\
  \widehat B & = & \frac1{2N_0}\sum^{N_0}_{k=1}\xi^B_k B_{4k}(xr_k), \\
  \widehat C & = & \frac1{2N_0}\sum^{N_0}_{k=1}\xi^C_k B_{4k}(xr_k),
\end{array}\right.
\end{eqnarray}
where $B_{4k}$ are the $4k$-order Boubaker polynomials
\cite{Tabatabaei}-\cite{Fridjine2}, $r_k$ are $B_{4k}$ minimal
positive roots, $N_0$ is a prefixed integer, and
$a_k|_{k=1,\ldots,N_0}$ are unknown pondering real coefficients.

As a first step, the coefficients, the coefficients
$\xi^A_k|_{k=1,\ldots,N_0}$ are determined through Falaleev
approximation [yy]:
\begin{eqnarray}
G_n\approx \frac1{\pi}\Br{\ln\Pa{n+\frac34} +\gamma + \ln16}
\end{eqnarray}
The BPES solution for $\widehat A$ is obtained by determining the
non-null set of coefficients $\widehat \xi^A_k|_{k=1,\ldots,N_0}$
that minimizes the absolute difference $\Delta_{N_0}$:
\begin{eqnarray}
\Delta_{N_0} =
\abs{\frac1{2N_0}\sum^{N_0}_{k=1}\widehat\xi^A_k\Lambda_k -
\frac1{\pi}(\gamma+\ln16)}
\end{eqnarray}
with
$$
\Lambda_k = \frac34 \int^1_0 \sum^{N_0}_{k=1} xB_{4k}(xr_k)dx.
$$

Values of $\widehat B$ and $\widehat C$ are consecutively deduced
from coefficients $\widehat \xi^B_k|_{k=1,\ldots,N_0}$ and $\widehat
\xi^C_k|_{k=1,\ldots,N_0}$ which minimize the absolute difference
$\Delta'_{N_0}$:
\begin{eqnarray}
\Delta'_{N_0} = \abs{\frac1{2N_0}\sum^{N_0}_{k=1}\widehat\xi^A_k X_k
- \frac1{\pi}(\gamma+\ln16) +
\frac1{2N_0}\sum^{N_0}_{k=1}\widehat\xi^B_k Y_k\Pa{1-
\frac1{2N_0}\sum^{N_0}_{k=1}\widehat\xi^C_k Z_k}}
\end{eqnarray}
with
$$
X_k = \frac3{4\pi}\int^1_0 \sum^{N_0}_{k=1}
xB_{4k}(xr_k)dx\quad\text{and}\quad
Y_k=Z_k=\int^1_0\sum^{N_0}_{k=1}B_{4k}(xr_k)dx.
$$
Hence the final solution is:
\begin{eqnarray}
G_n = \frac1{\pi}\ln(n+A) + \frac1{\pi}(\gamma+\ln16) -
\frac{B}{n+C}
\end{eqnarray}
with
$$
\left\{\begin{array}{ccc}
A =\widehat A& = & \frac1{2N_0}\sum^{N_0}_{k=1}\xi^A_k B_{4k}(xr_k), \\
B =\widehat B& = & \frac1{2N_0}\sum^{N_0}_{k=1}\xi^B_k B_{4k}(xr_k), \\
C =\widehat C& = & \frac1{2N_0}\sum^{N_0}_{k=1}\xi^C_k B_{4k}(xr_k).
\end{array}\right.
$$

\section{Results, plots and discussion}

Numerical solutions obtained by the given method are gathered in
Table~\ref{tab:1} along with precedent refereed approximations by
Falaleev \cite{Falaleev} and Brutman \cite{Brutman}.
\begin{table}
 {\scriptsize\centering\caption{\label{tab:1} Landau constants values.}
 \begin{tabular}{|c|c|c|c|c|c|c|c|c|c|}
  \hline
  \multirow{2}{*}n & \multicolumn{3}{|c|}{Landau constants} & \multicolumn{2}{|c|}{Quadratic Error vs. Exact} \\
  \cline {2-6}&Falaleev Approximation \cite{Falaleev} & Brutman Approximation \cite{Brutman}
  & BPES Approximation & Ref.~\cite{Falaleev} & Ref.~\cite{Brutman} \\
  \hline
  0 & 0.97463516 & 0.97469795 & 0.97473411 & 9.7924E-9 & 1.30765E-9 \\
  \hline
  1 & 1.24433844 & 1.24440124 & 1.24449240 & 2.37026E-8 & 8.31042E-9 \\
  \hline
  2 & 1.38820978 & 1.38827257 & 1.38841873 & 4.36628E-8 & 2.13632E-8 \\
  \hline
  3 & 1.48693516 & 1.48699795 & 1.48719911 & 6.96731E-8 & 4.04659E-8 \\
  \hline
  4 & 1.56218004 & 1.56224284 & 1.56249900 & 1.01733E-7 & 6.56187E-8 \\
  \hline
  5 & 1.62299481 & 1.62305761 & 1.62336877 & 1.39843E-7 & 9.68215E-8 \\
  \hline
  6 & 1.67403346 & 1.67409626 & 1.67446242 & 1.84004E-7 & 1.34074E-7 \\
  \hline
  7 & 1.71800808 & 1.71807088 & 1.71849204 & 2.34214E-7 & 1.77377E-7 \\
  \hline
  8 & 1.75663844 & 1.75670124 & 1.75717740 & 2.90474E-7 & 2.2673E-7 \\
  \hline
  9 & 1.79108390 & 1.79114669 & 1.79167785 & 3.52784E-7 & 2.82133E-7 \\
  \hline
  10 & 1.82216319 & 1.82222598 & 1.82281214 & 4.21145E-7 & 3.43585E-7 \\
  \hline
  11 & 1.85047605 & 1.85053884 & 1.85118001 & 4.95555E-7 & 4.11088E-7 \\
  \hline
  12 & 1.87647497 & 1.87653777 & 1.87723393 & 5.76015E-7 & 4.84641E-7 \\
  \hline
  13 & 1.90050977 & 1.90057257 & 1.90132373 & 6.62525E-7 & 5.64244E-7 \\
  \hline
  14 & 1.92285648 & 1.92291928 & 1.92372544 & 7.55085E-7 & 6.49896E-7 \\
  \hline
  15 & 1.94373675 & 1.94379954 & 1.94466070 & 8.53696E-7 & 7.41599E-7\\
  \hline
  16 & 1.96333123 & 1.96339403 & 1.96431019 & 9.58356E-7 & 8.39352E-7 \\
  \hline
  17 & 1.98178915 & 1.98185195 & 1.98282311 & 1.06907E-6 & 9.43155E-7 \\
  \hline
  18 & 1.99923515 & 1.99929795 & 2.00032411 & 1.18583E-6 & 1.05301E-6 \\
  \hline
  19 & 2.01577445 & 2.01583725 & 2.01691841 & 1.30864E-6 & 1.16891E-6
  \\
  \hline
 \end{tabular}}
\end{table}

\begin{figure}[htbp]
\centering
\includegraphics[height=3in,width=5in]{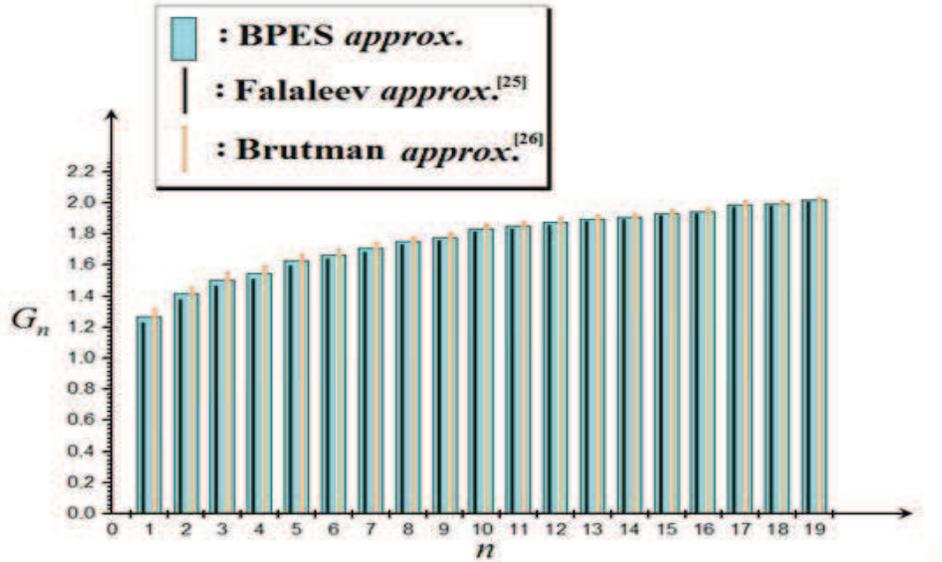}\caption{Solution
plots}\label{fig:1}
\end{figure}

Plots of the BPES solution are presented in Fig.~\ref{fig:1}, along
with referred solutions \cite{Falaleev,Brutman}.

\begin{figure}[htbp]
\centering
\includegraphics[height=3in,width=4.5in]{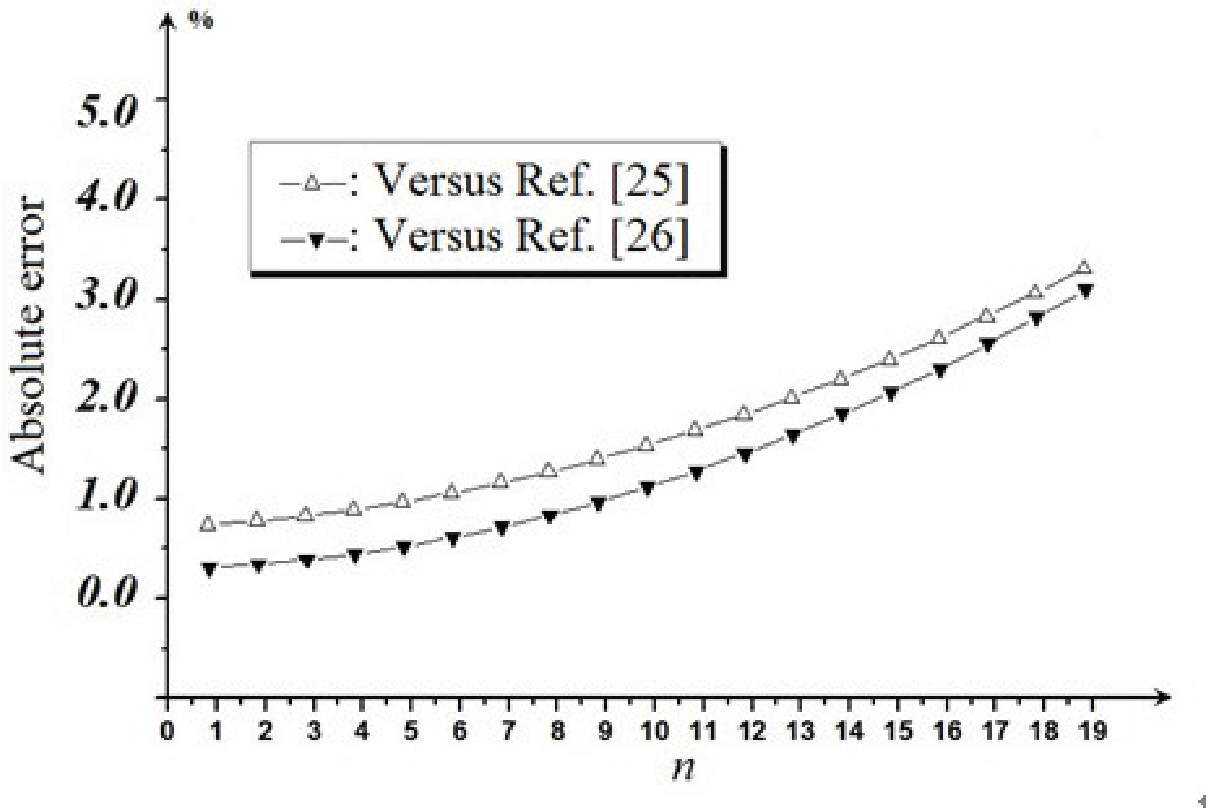}\caption{Absolute  error
plots}\label{fig:2}
\end{figure}

Figure~\ref{fig:2} displays the obtained values along with those
recorded by Falaleev \cite{Falaleev} and Brutman \cite{Brutman}. For
accuracy purposes, error analysis has been carried out for the two
referred datasets. Examination of the quadratic error plots
(Fig.~\ref{fig:2}) shows that the amplitudes of the error are more
exaggerated according to Falaleev approximation \cite{Falaleev},
particularly for low values of $n$.

Moreover, Figure~\ref{fig:2} monitors an obvious logarithmic profile
in the range with a quadrature error which doesn't exceed 3.5\% on
the whole range (Fig.~\ref{fig:2}). The obtained profile is in good
agreement with the values recorded elsewhere
\cite{Falaleev}-\cite{Stanley}.

\section{Conclusion}

In this study, approximation formulas for evaluating Landau
constants have been presented and discussed.  The  proposed estimate
has been  compared  with  two  other estimates  which  are of
special importance in approximation  theory. Results have been
favorable for the performed method in terms of both convergence and
accuracy.



\end{document}